\documentclass{Interspeech2024}
\usepackage{graphicx}
\usepackage{amsmath,amssymb}
\usepackage{nicefrac}
\usepackage{tikz}
\usepackage{multirow}
\usepackage{makecell}
\usepackage[shortcuts]{extdash}
\usepackage{xspace}
\usepackage{placeins}
\usepackage[super]{nth}

\usepackage{pifont}
\newcommand{\cmark}{\ding{51}}%
\newcommand{\xmark}{\ding{55}}%

\interspeechcameraready

\title{Outlier Reduction with Gated Attention for Improved Post-training Quantization in Large Sequence-to-sequence Speech Foundation Models}

\name[affiliation={1}]{Dominik}{Wagner}
\name[affiliation={1}]{Ilja}{Baumann}
\name[affiliation={1}]{Korbinian}{Riedhammer}
\name[affiliation={1,2}]{Tobias}{Bocklet}

\address{
  $^1$Technische Hochschule Nürnberg Georg Simon Ohm \\
  $^2$Intel Labs}
\email{dominik.wagner@th-nuernberg.de}


\definecolor{nodefill}{RGB}{218,232,252}
\definecolor{nodedraw}{RGB}{108,142,191}

\let\OLDthebibliography\thebibliography
\renewcommand\thebibliography[1]{
  \OLDthebibliography{#1}
  \setlength{\parskip}{0pt}
  \setlength{\itemsep}{0pt plus 0.3ex}
}

\begin{document}

\maketitle

\begin{abstract}
This paper explores the improvement of post-training quantization (PTQ) after knowledge distillation in the Whisper speech foundation model family. 
We address the challenge of outliers in weights and activation tensors, known to impede quantization quality in transformer-based language and vision models. 
Extending this observation to Whisper, we demonstrate that these outliers are also present when transformer-based models are trained to perform automatic speech recognition, necessitating mitigation strategies for PTQ. 
We show that outliers can be reduced by a recently proposed gating mechanism in the attention blocks of the student model, enabling effective 8-bit quantization, and lower word error rates compared to student models without the gating mechanism in place. 
\end{abstract}

\section{Introduction}
Foundation models are deep neural networks trained on extensive and diverse datasets, capable of addressing a wide range of downstream tasks with minimal or no adaptation required. 
Speech foundation models (SFMs) like wav2vec 2.0 \cite{baevski2020w2v2}, HuBERT \cite{weining2021hubert}, and Whisper \cite{radford2022whisper} have demonstrated remarkable performance across various speech-related tasks. 
However, improved performance also led to a notable increase in both the number of parameters and computational complexity. 
The increased demand for computational resources not only results in higher energy consumption but also limits the accessibility of SFM-based applications on resource-constrained devices. 
Consequently, there is a growing focus on enhancing the efficiency of SFMs \cite{gandhi2023distilwhisper,peng2024owsmctc,shao2023whisperkdq,fish23_interspeech,chang2022distilhubert}.

Common approaches to reduce the size of automatic speech recognition (ASR) systems include knowledge distillation (KD) \cite{chang2022distilhubert,gong22_interspeech,ashihara22_interspeech}, model quantization \cite{yeh2022efficient,ding22c_interspeech,zhen22_interspeech,fasoli21_interspeech,fish23_interspeech}, weight pruning \cite{takafumi2020distill,lodagala2022pada,cheng2021parp}, or combinations thereof \cite{kim21m_interspeech,ding2024usmlite}. 
Two widely used quantization techniques are quantization-aware training (QAT) and post-training quantization (PTQ), with the later being the focus of this study.
QAT involves simulating quantization operations during model training. 
PTQ methods are typically easier to apply, as they either require no direct interaction with the model or involve passing only a small calibration dataset through the model to determine the optimal quantization parameters. 

Prior research has shown that weight quantization has minimal effect on the accuracy of transformer-based systems \cite{bondarenko-etal-2021-understanding,dettmers2022gptint}, whereas activation tensors exhibit significantly wider value ranges  \cite{yao2022zeroquant,xiao2023smoothquant}, making them more difficult to quantize. 

Several efforts have been made to reduce the size of the Whisper SFM, a model also utilized in this study.  
These approaches either focus exclusively on KD \cite{gandhi2023distilwhisper,ferraz2024multilingual}, QAT \cite{shao2023whisperkdq}, or apply PTQ to calibrate personalized quantization schemes for particular speakers \cite{fish23_interspeech}. 

Analyzing the intricacies of SFMs is crucial for optimizing their performance, especially in resource-constrained environments. 
By understanding why models perform less effectively after quantization, we may gain insights into the determinants of their limitations. 

Investigations into the behavior of transformer-based language models indicate that these models learn outliers within their weights and activation tensors \cite{luo2021positional,wei2022outlier,dettmers2022gptint,kovaleva-etal-2021-bert,bondarenko-etal-2021-understanding,bondarenko2023quantizable,xiao2023smoothquant}. 
Outliers typically appear within a limited, fixed set of hidden dimensions but materialize across various layers irrespective of the input sequence. 
Moreover, these outliers influence the quality of the model's predictions, and attempts to mitigate their impact, e.g., by dropping the corresponding values, can result in a significant degradation of the model's performance \cite{kovaleva-etal-2021-bert}. 
The presence of outliers within hidden dimensions also poses challenges for model quantization due to the trade-off between rounding errors and clipping errors \cite{dettmers2022gptint,bondarenko-etal-2021-understanding,bondarenko2023quantizable}. 
Previous works argue that these anomalies stem from specific behaviors exhibited by attention heads attempting to either learn a null operation or a partial update of the residuals \cite{kovaleva-etal-2019-revealing,bondarenko-etal-2021-understanding,bondarenko2023quantizable}. 
To obtain the zero values necessary for a non-update scenario in the attention matrix, the softmax input undergoes continual amplification during training, resulting in outliers in other network components. 

So far, investigations into outliers have been focused on transformer-based language and vision models. 
In this work, we show that the observations made in the text and image domain also translate to the speech domain and that quantizing both weight and activation tensors to 8-bit benefits from outlier mitigation. 
In particular, we employ the gating mechanism for attention blocks introduced in \cite{bondarenko2023quantizable}, and demonstrate that Whisper-based models distilled with this gating mechanism in place learn smaller outliers and exhibit less performance degradation after quantization. 

Knowledge distillation provides a robust framework for leveraging the knowledge encapsulated within the pretrained Whisper model, while also providing freedom in the choice of the student architecture. 
Given the unavailability of the original training data and the significant computational resources required for Whisper pretraining, we opt for student-teacher training using a diverse dataset of $\sim$16k hours of English speech collected from various publicly available sources comprising a large number of speakers and speaking styles to transfer knowledge to the student.
Our focus is on analyzing outliers and evaluating the effectiveness of gated attention in mitigating them, along with improving ASR performance after PTQ. 
\newline
\newline
Our main contributions are:
\begin{itemize}
    \item Identification of outlier behavior in Whisper, mirroring findings in transformer-based language and vision models
    \item Application of a gated attention mechanism to effectively address outliers in hidden dimensions of SFMs
    \item Creation of several smaller distilled versions of Whisper using student-teacher training to reduce the size of both encoder and decoder
    \item Demonstration of a practical framework for post-training quantization of SFMs to INT8, enhancing ASR efficiency and deployability
\end{itemize}

\section{Method}
\subsection{Knowledge distillation}\label{ssec:kd}
Knowledge distillation (KD) \cite{hinton15distill} is a learning framework that entails training a more compact student model to emulate the performance of a larger teacher model. 
KD involves training by aligning the student's predictions with those of the teacher. 
Following \cite{kim-rush-2016-sequence}, we utilize a linear combination of cross-entropy loss across a set of $N$ target labels $\boldsymbol{y}_{1:N} = \lbrace y_1 \ldots , y_N \rbrace$ and the Kullback-Leibler ($KL$) divergence \cite{kullback1955kld} to optimize the student:
\begin{equation}
\mathcal{L}= \alpha_{\text{CE}}  \left( - \sum_{i=1}^N p(y_i \mid \boldsymbol{y}_{<i}, \boldsymbol{H}_{1:M})  \right)
+  \alpha_{\text{KL}} \left( KL\left(\mathbf{S}, \mathbf{T} \right) \right),
\end{equation}
where $\mathbf{S}$ and $\mathbf{T}$ denote the output probability distribution of the student and the teacher model, respectively. 
The constant weighting factors are set to $\alpha_{\text{CE}}=1$ and $\alpha_{\text{KL}}=0.8$ based on \cite{gandhi2023distilwhisper}.
\subsection{Post-training quantization}
We consider post-training quantization (PTQ), where a trained full precision (FP32) model is converted into an 8-bit (INT8) fixed-point model directly without any additional training. 
Quantization is the process of mapping the values of a tensor $\mathbf{x} \in \mathbb{R}$ to corresponding values on an integer grid $\mathbf{x}_q \in \mathbb{Z}$. 
Following \cite{xiao2023smoothquant,bondarenko2023quantizable}, we emulate the quantization process according to \cite{Jacob_2018_CVPR}:
\begin{equation}
\mathbf{x}_q = s \cdot \left( \operatorname{min}\left(  \operatorname{max}\left( \left\lfloor\frac{\mathbf{x}}{s}\right\rceil + z , 0 \right), 2^b-1 \right) - z \right),
\end{equation}
where $\mathbf{x}$ represents either model weights or activation tensors, $s \in \mathbb{R}_{+}$ is a scaling factor specifying the quantization step size, $b \in \mathbb{N}$ is the target bitwidth, $z \in \mathbb{Z}$ is the zero-point, and $\lfloor\cdot\rceil$ indicates rounding to the nearest integer. 

The quantization procedure in our experiments encompasses both weight and activation quantization to INT8. 
Since activations are dependent on the input data, a critical aspect of the PTQ process for activation tensors involves identifying appropriate minimum and maximum values for each quantizer (i.e., the scaling factor $s$ applied to the full-precision values). 
Several methods are available for establishing the boundaries of the interval. 
We employ a static approach, which estimates the quantization range based on 16 batches from the validation set utilizing an exponential moving average of the minimum and maximum values across those batches \cite{krishnamoorthi2018quantizing}. 
For weight quantization, we use the full range of the weight tensors. 

All weights in every layer of the student's encoder and decoder blocks are quantized. 
Despite the presence of gating mechanisms in each encoder layer, we found that the last encoder layer retains a relatively large dynamic range, resulting in imprecise INT8 representation and consequently higher word error rates. 
Therefore, for activation quantization, all layers of the student model, except for the final output projection in the last encoder layer along with the final layer norm, are quantized. 
\subsection{Gated attention}
\begin{figure*}[t]
  \includegraphics[width=\textwidth]{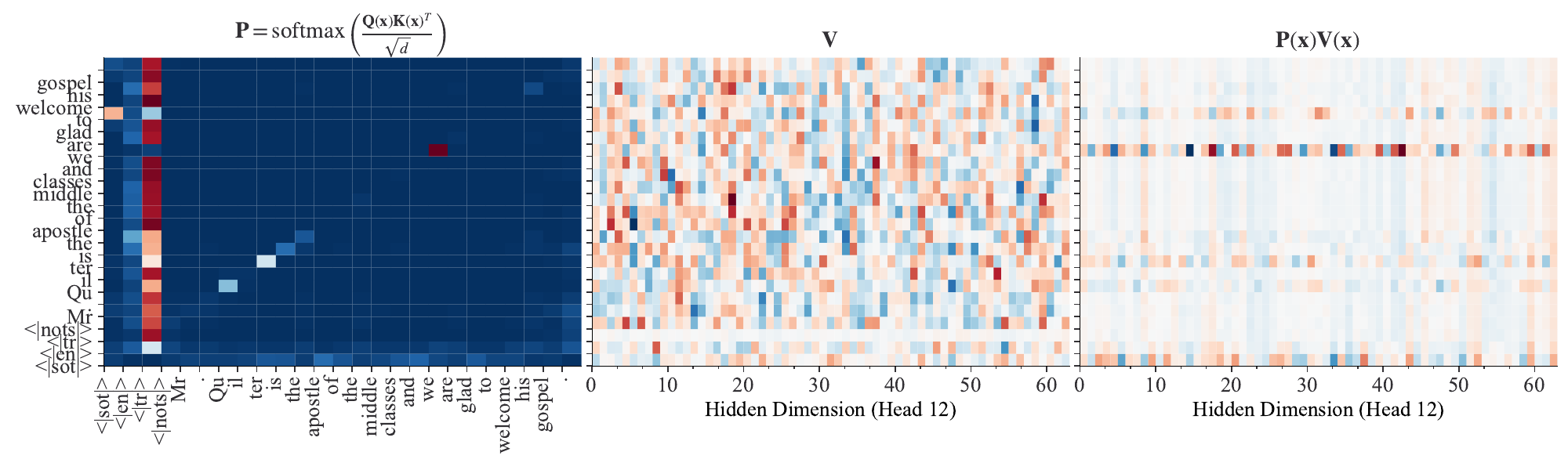}
  \vspace{-8.5mm}
  \caption{Behavior of the self-attention mechanism in the pretrained Whisper model (\texttt{whisper-large-v2}) computed for the first example of the LibriSpeech test-clean set. The left matrix shows the attention probabilities $\mathbf{P}$, where $d=64$ is the dimensionality of the attention head. The middle matrix are the values $\mathbf{V}$ in the twelfth attention head. The right matrix is the product of the two.}
  \label{fig:outlier}
  \vspace{-6mm}
\end{figure*}
A recently proposed conditional gating method that helps controlling the update process of hidden representations has been shown to be effective for reducing outliers in transformer-based language and vision models \cite{bondarenko2023quantizable}. 
This approach allows the model to selectively retain or nullify updates to the representation of specific tokens, independent of the attention probabilities and values. 

In \cite{bondarenko2023quantizable}, a gating function $\mathcal{G}$ is activated through a sigmoid nonlinearity $\sigma$ and subsequently multiplied with the attention outputs using the Hadamard product:
\begin{equation}\label{eq:self_att}
\mathcal{A}(\mathbf{x}) := \operatorname{softmax}\left(\frac{\boldsymbol{Q}(\mathbf{x}) \boldsymbol{K}(\mathbf{x})^T}{\sqrt{d}}\right) \boldsymbol{V}(\mathbf{x})
\end{equation}
\begin{equation}\label{eq:gated_att}
\operatorname{gated\_att}(\mathbf{x}) := \sigma(\mathcal{G}(\mathbf{x})) \odot \mathcal{A}(\mathbf{x})
\end{equation}
$\mathcal{A}(\mathbf{x})$ is the self-attention mechanism defined in \cite{vaswani17attention} with the trainable linear projections $\boldsymbol{Q}$, $\boldsymbol{K}$ and $\boldsymbol{V}$, as well as an input $\mathbf{x}$. 
Whisper employs multi-headed self-attention, in which the feature representations are divided into $n$ parts of dimensionality $d$.
The attention mechanism is applied to each $n$ and concatenated to the final output. 
$\mathcal{G}$ is a neural network with a single linear layer, trained along with the rest of the model. 
We substitute Equation~\ref{eq:self_att} with Equation~\ref{eq:gated_att} across all encoder and decoder layers of the student model and use a single $\mathcal{G}$ that is shared across different attention heads but not across different token positions. 
The authors of \cite{bondarenko2023quantizable} use one gating function per attention head that is shared across different token position in their main experiments. 
They also proposed an additional method for outlier mitigation called clipped softmax. 
We also tried these methods in preliminary investigations but found that the gated attention mechanism was more effective than clipped softmax and gating on a per-head basis did perform worse than sharing the gate across attention heads. 
\subsection{Whisper}
Whisper \cite{radford2022whisper} is a family of transformer-based sequence-to-sequence \cite{vaswani17attention} SFMs trained to perform multiple tasks such as multilingual ASR, language identification, and speech translation.
The models are trained on $\sim$680k hours of proprietary data retrieved from the world wide web and are available in five sizes ranging from 39M parameters to 1.55B parameters. 
All Whisper SFMs utilize an encoder-decoder structure but differ in parameters such as the number of transformer blocks, the number of attention heads, and hidden layer dimensions. 

The Whisper encoder $\mathcal{E}$ maps a sequence of $L$ log-Mel spectrogram features $\boldsymbol{F}$ obtained from the raw audio waveform $\boldsymbol{A}$: $\boldsymbol{F}(\boldsymbol{A})_{1:L} = \lbrace \boldsymbol{f}_1, \ldots , \boldsymbol{f}_L \rbrace$ to a sequence of $M$ hidden representations $\boldsymbol{H}_{1:M}$:
\begin{equation*}
   \mathcal{E} : \boldsymbol{F}(\boldsymbol{A})_{1:L} \mapsto \boldsymbol{H}_{1:M}. 
\end{equation*}
The decoder predicts the probabilities for the next token $y_i$, based on the preceding tokens tokens $\boldsymbol{y}_{<i}$ and the hidden representations $\boldsymbol{H}_{1:M}$: $p(y_i \mid \boldsymbol{y}_{<i}, \boldsymbol{H}_{1:M})$. 
The model is trained on pairs of log-Mel
spectrogram features and target transcriptions, using the cross-entropy objective.

We distill the 1.55B parameter version of Whisper (\texttt{whisper-large-v2}) into three variants using either 8, 16 or 24 encoder layers instead of the original 32 layers. 
Based on findings that drastically reducing or even eliminating the Whisper decoder does not significantly compromise performance, we reduce its size to 2 layers \cite{peng2024owsmctc,gandhi2023distilwhisper}. 
The total number of trainable parameters for the resulting student model variants are 283M, 440M, and 598M parameters, respectively.


\section{Experiments and Results}
\subsection{Data}
We used a combination of four publicly available English datasets as our training corpus: People's Speech \cite{galvez2021the}, CommonVoice (version 16) \cite{ardila2020cv}, LibriSpeech \cite{panayotov2015librispeech}, and Voxpopuli \cite{wang2021voxpopuli}. 
The combined training data comprises approximately $\sim$16k hours of speech sourced from diverse domains such as audiobooks, political speeches, interviews, and narrated Wikipedia articles. 
In addition to evaluating ASR performance on in-distribution (ID) data comprised of the test portions of the training corpora, we employed three out-of-distribution (OOD) test sets to further asses the robustness of the gated attention approach towards PTQ: TED-LIUM \cite{rousseau2012ted}, Fleurs \cite{conneau2022fleurs}, and Gigaspeech \cite{chen21o_interspeech}. 
These datasets encompass diverse speech characteristics and additional domains such as podcasts and talks. 
\subsection{Evaluation metrics}
To assess ASR performance, we examined word error rates (WERs) across both ID and OOD test sets.
We evaluated ASR performance for differently sized student models with and without INT8 quantization on weights and activations using either coventional attention or the gating mechanism. 
Following \cite{bondarenko2023quantizable}, we also analyzed outliers measured by kurtosis and infinity norm $\left \lVert \cdot \right \rVert_{\infty}$, exploring the impact of gated attention across different model sizes. 
Kurtosis was averaged across the outputs of all attention layers. 
Based on \cite{bondarenko-etal-2021-understanding}, we count values that exceed six standard
deviations from the mean of the activation tensor as outliers.
\subsection{Modeling and architecture details}
Leveraging the shared dimensionality of hidden layers between teacher and student networks, we initialized each student using the pre-trained weights of the teacher \cite{sanh2020distilbert,shleifer2020pretrained,gandhi2023distilwhisper}. 
In particular, we selectively copied the weights of 8, 16 or 24 layers from the teacher encoder. 
For example, when the number of layers in the encoder was reduced by a factor of 4 (i.e., 8 layers instead of 32), we copied every \nth{4} weight matrix from the teacher starting with the first one. 
In the student decoder, we copied the weights from the initial and final layers of the teacher. 

Each model was trained for $2 \times 10^5$ steps with an effective batch size of 64, which amounts to two training epochs. 
We used the AdamW optimizer \cite{loshchilov2018decoupled} ($\lambda = 10^{-4}$, $\epsilon = 10^{-8}$, $\beta_1 = 0.99$, $\beta_2 = 0.999$) with an initial learning rate of $10^{-4}$, a linear schedule and a warm-up phase of 1000 steps. 
The models used in the evaluation were selected based on the lowest WER achieved on the validation set comprised of all validation portions from the four datasets used for training. 
We utilized greedy decoding for its increased inference speed. 
\subsection{Whisper exhibits outlier behavior similar to LMs}
As an initial analysis, we followed \cite{bondarenko2023quantizable} and visualized the behavior of the multi-headed self-attention mechanism in the Whisper decoder. 
The large version of Whisper employs 20 attention heads with 64 hidden dimensions each.  
Figure~\ref{fig:outlier} illustrates the attention mechanism in the \nth{31} layer in the decoder part of the pretrained 1.55B parameter version of Whisper.  
We observe that the attention head predominantly allocates its probability mass to the transcription token \texttt{<|tr|>}, while the same token has small values associated with it in $\mathbf{V}$ (cf. third row from the bottom in the middle matrix). 
As a result, the product between the two is small, representing only a minimal or no update of the hidden representation. 
Only a small portion of the probability mass is distributed to other tokens (in this case mostly the token \texttt{we}), resulting in a local update of the hidden representation for those tokens.
Similar patterns can be found across all decoder layers and attention heads. 
These results are in line with the findings on transformer-based language and vision models \cite{clark-etal-2019-bert,bondarenko2023quantizable}, supplementing them with SFMs. 
\subsection{Gated attention improves PTQ in student models}
\begin{figure}[t]
  \includegraphics[width=\linewidth]{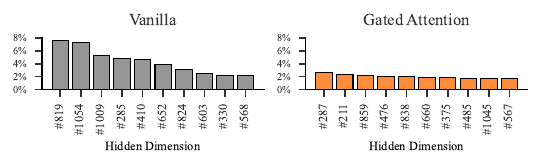}
  \caption{Top 10 shares of activation outliers per hidden dimension at the output projection of the self-attention block of the last decoder layer in trained student models. Left are the relative outliers for a student trained without gated attention and right are the relative outliers with gated attention. Both models were trained using 24 layers for the encoder. The hidden dimensions are zero-indexed. Each output projection layer has a dimensionality of 1280.}
  \label{fig:outlier_dim}
  \vspace{-2mm}
\end{figure}
Figure~\ref{fig:outlier_dim} illustrates which hidden dimensions contribute the most outliers for one student model trained without gated attention (left) and another one trained with gated attention (right). 
Outliers were counted separately for each dimension in the output projection of the self-attention block in the last decoder layer (i.e., the \nth{2} decoder layer). 
We see that the distribution of outliers becomes more uniform and each dimension contributes less to the overall outliers with the gating mechanism in place. 
For example, with conventional attention, the hidden dimensions \#819 and \#1054 contribute $\sim$15\% of all outliers, whereas with gated attention the two highest shares (dimension \#287 and \#211) amount only to a total of $\sim$5\% relative to all outliers. 
\begin{table}[t]
\caption{Average kurtosis and maximum infinity norm on the full ID test set for different model sizes before quantization.}
\label{tab:outlier_exp}
\setlength{\tabcolsep}{10pt}
\centering
\begin{tabular}{ccrr}
\toprule
\multirow{2}{*}{\makecell{\textbf{Decoder} \\ \textbf{Layers}}} & \multirow{2}{*}{\makecell{\textbf{Gated} \\ \textbf{Attention}}} & 
\multirow{2}{*}{\makecell{\textbf{Average} \\ \textbf{Kurtosis}}} & 
\multirow{2}{*}{\makecell{\textbf{Maximum} \\ \textbf{Inf. Norm}}} \\
&  &  &  \\
\midrule
\multirow[c]{2}{*}{24} & \xmark & 105.4 & 66.8 \\
 & \cmark & 42.8 & 44.0 \\
\cline{1-4}
\multirow[c]{2}{*}{16} & \xmark & 48.5 & 35.3 \\
 & \cmark & 19.5 & 29.8 \\
\cline{1-4}
\multirow[c]{2}{*}{8} & \xmark & 28.6 & 29.9 \\
 & \cmark & 20.7 & 23.7 \\
\cline{1-4}
\bottomrule
\end{tabular}
\vspace{-6mm}
\end{table}


Table~\ref{tab:outlier_exp} shows the average kurtosis and $\left \lVert \cdot \right \rVert_{\infty}$ across test sets with and without gated attention. 
Consistently lower outlier metrics were observed for models trained with gated attention compared to those trained without gated attention, indicating improved robustness and stability in the former case.
\begin{table}[t]
\caption{Word Error Rates on different in-distribution (ID) and out-of-distribution (OOD) test sets.}
\label{wer_exp}
\setlength{\tabcolsep}{4.5pt}
\centering
\begin{tabular}{cccrrr}
\toprule
\multirow[c]{2}{*}{\textbf{Dataset}} & 
\multirow[c]{2}{*}{\makecell{\textbf{INT8} \\ \textbf{Quant.}}} & 
\multirow[c]{2}{*}{\makecell{\textbf{Gated} \\ \textbf{Attention}}}  & \multicolumn{3}{c}{\textbf{WER}} \\
 &  & & \makecell{24 \\ \footnotesize{layer}} & \makecell{16 \\ \footnotesize{layer}}  & \makecell{8 \\ \footnotesize{layer}} \\
\midrule
\multirow[c]{4}{*}{\makecell{Voxpopuli \\ (ID)}} & \multirow[c]{2}{*}{\xmark} & \xmark & 13.7 & 15.6 & 19.2 \\
 &  & \cmark & 12.6 & 13.8 & 18.1 \\
\cline{2-6}
 & \multirow[c]{2}{*}{\cmark} & \xmark & 21.7 & 22.2 & 24.1 \\
 &  & \cmark & 13.9 & 15.6 & 19.6 \\
\cline{1-6} \cline{2-6}
\multirow[c]{4}{*}{\makecell{LibriSpeech \\ test-clean \\ (ID)}} & \multirow[c]{2}{*}{\xmark} & \xmark & 7.0 & 6.7 & 10.7 \\
 &  & \cmark & 6.3 & 6.9 & 10.0 \\
\cline{2-6}
 & \multirow[c]{2}{*}{\cmark} & \xmark  & 9.7 & 12.7 & 13.7 \\
 &  & \cmark & 7.7 & 8.6 & 12.3 \\
\cline{1-6} \cline{2-6}
\multirow[c]{4}{*}{\makecell{LibriSpeech \\ test-other \\ (ID)}} & \multirow[c]{2}{*}{\xmark} & \xmark & 13.1 & 14.6 & 21.1 \\
 &  & \cmark & 12.7 & 14.0 & 20.6 \\
\cline{2-6}
 & \multirow[c]{2}{*}{\cmark} & \xmark & 18.4 & 23.2 & 24.0 \\
 &  & \cmark & 14.4 & 15.8 & 22.8 \\
\cline{1-6} \cline{2-6}
\multirow[c]{4}{*}{\makecell{CommonVoice 16 \\ (ID)}} & \multirow[c]{2}{*}{\xmark} & \xmark & 23.6 & 27.6 & 36.5 \\
 &  & \cmark & 22.4 & 26.2 & 36.3 \\
\cline{2-6}
 & \multirow[c]{2}{*}{\cmark} & \xmark & 34.4 & 33.2 & 39.3 \\
 &  & \cmark & 25.0 & 28.4 & 38.6 \\
\cline{1-6} \cline{2-6}
\multirow[c]{4}{*}{\makecell{People's Speech \\ (ID)}} & \multirow[c]{2}{*}{\xmark} & \xmark & 33.7 & 33.9 & 46.1 \\
 &  & \cmark & 32.7 & 35.0 & 41.6 \\
\cline{2-6}
 & \multirow[c]{2}{*}{\cmark} & \xmark & 38.7 & 45.6 & 47.6 \\
 &  & \cmark & 35.5 & 37.6 & 44.7 \\
\cline{1-6} \cline{2-6}
\multirow[c]{4}{*}{\makecell{TED-LIUM \\ (OOD)}} & \multirow[c]{2}{*}{\xmark} & \xmark & 13.6 & 15.0 & 19.6 \\
 &  & \cmark & 14.3 & 14.9 & 18.0 \\
\cline{2-6}
 & \multirow[c]{2}{*}{\cmark} & \xmark & 19.7 & 22.4 & 27.1 \\
 &  & \cmark & 15.5 & 17.2 & 19.8 \\
\cline{1-6} \cline{2-6}
\multirow[c]{4}{*}{\makecell{Gigaspeech \\ (OOD)}} & \multirow[c]{2}{*}{\xmark} & \xmark & 18.4 & 19.9 & 26.2 \\
 &  & \cmark & 17.9 & 20.0 & 25.7 \\
\cline{2-6}
 & \multirow[c]{2}{*}{\cmark} & \xmark & 23.1 & 25.7 & 28.5 \\
 &  & \cmark & 19.2 & 21.5 & 27.8 \\
\cline{1-6} \cline{2-6}
\multirow[c]{4}{*}{\makecell{Fleurs \\ (OOD)}} & \multirow[c]{2}{*}{\xmark} & \xmark & 17.1 & 17.7 & 25.2 \\
 &  & \cmark & 15.8 & 18.6 & 25.5 \\
\cline{2-6}
 & \multirow[c]{2}{*}{\cmark} & \xmark & 25.2 & 21.3 & 28.9 \\
 &  & \cmark & 17.4 & 20.5 & 28.1 \\
\cline{1-6} \cline{2-6}
\bottomrule
\end{tabular}
\vspace{-6mm}
\end{table}

Table~\ref{wer_exp} presents WERs on each ID and OOD test set for student models with 24, 16, and 8 encoder layers before and after INT8 quantization of weights and activations. 
Each model was trained once with the gated attention mechanism in place and once without. 
The 24-layer encoder system demonstrated the best overall performance across the test sets and WERs increased with fewer encoder layers used in the student model. 
Before quantization, using the gated attention mechanism during student-teacher training resulted in similar WERs compared to training without gated attention across all test sets. 
However, a notable difference emerges after quantization. 
With gated attention, Table~\ref{wer_exp} shows only slight increases in WERs relative to non-quantized models, whereas quantized models trained without gated attention experience significant increases in WERs across all test sets. 
This highlights the effectiveness of gated attention in maintaining ASR performance after quantization.
Additionally, in some cases, the gating mechanism also improves the full precision performance. 
However, this is not consistent across all test sets and encoder layer configurations. 

Our findings align with outlier analyses conducted for language and vision models \cite{kovaleva-etal-2021-bert,bondarenko-etal-2021-understanding,wei2022outlier,bondarenko2023quantizable}, showing that the importance of outlier mitigation is crucial to control performance degradation in PTQ. 
\vspace{-4mm}
\section{Conclusions}
\vspace{-1mm}
In line with observations in transformer-based language and vision models, we found that outliers generated by the attention mechanism, attempting to perform no-update operations, also manifest in the Whisper model architecture. 
These outliers pose challenges when applying post-training quantization, particularly concerning activation quantization. 
We focused on reducing the model size through student-teacher training, followed by applying INT8 quantization to both weight and activation tensors. 
To enhance robustness against outliers, we used a gated attention mechanism that learns to perform the null operation when necessary. 
Our experiments demonstrated that the student model equipped with the gating mechanism achieved similar WERs as the model without gated attention at floating-point precision. 
However, when quantization was applied, models with gated attention exhibited significantly more stable WERs. This underscores the efficacy of gated attention in complementing student-teacher training, particularly in scenarios where the ultimate goal involves weight and activation quantization. 
\vspace{-1mm}
\section{Acknowledgments}
\vspace{-1mm}
We gratefully acknowledge the scientific support and HPC resources provided by the Erlangen National High Performance Computing Center (NHR@FAU) of the Friedrich-Alexander-Universität Erlangen-Nürnberg (FAU) under the NHR project b196ac14. NHR funding is provided by federal and Bavarian state authorities. NHR@FAU hardware is partially funded by the German Research Foundation (DFG) – 440719683.
This work was supported by the Bavarian State Ministry of Science and the Arts under grant H.2-F1116.NÜ/61/2.

\newpage
\bibliographystyle{IEEEtran}
{\footnotesize{\bibliography{refs}}

\end{document}